\documentclass[conference]{IEEEtran}
\IEEEoverridecommandlockouts
\usepackage{cite}
\usepackage{amsmath,amssymb,amsfonts}
\usepackage{algorithmic}
\usepackage{graphicx}
\usepackage{textcomp}
\usepackage{xcolor}
\usepackage{subcaption}
\usepackage{url}
\ifCLASSOPTIONcompsoc
    \usepackage[caption=false, font=normalsize, labelfont=sf, textfont=sf]{subfig}
\else
\usepackage[caption=false, font=footnotesize]{subfig}
\fi
\usepackage[inline]{enumitem}
\def\BibTeX{{\rm B\kern-.05em{\sc i\kern-.025em b}\kern-.08em
    T\kern-.1667em\lower.7ex\hbox{E}\kern-.125emX}}
\begin{document}

\title{Input-Based Ensemble-Learning Method for Dynamic Memory Configuration of Serverless Computing Functions
}

\author{\IEEEauthorblockN{Siddharth Agarwal, Maria A. Rodriguez and Rajkumar Buyya}
\IEEEauthorblockA{\textit{Cloud Computing and Distributed Systems (CLOUDS) Laboratory} \\
\textit{School of Computing and Information Systems}\\
\textit{The University of Melbourne, Australia}\\
Email: siddhartha@student.unimelb.edu.au, \{maria.read, rbuyya\}@unimelb.edu.au}
}

\maketitle

\begin{abstract}
In today's Function-as-a-Service offerings, a programmer is usually responsible for configuring function memory for its successful execution, which allocates proportional function resources such as CPU and network. However, right-sizing the function memory force developers to speculate performance and make ad-hoc configuration decisions. Recent research has highlighted that a function's input characteristics, such as input size, type and number of inputs, significantly impact its resource demand, run-time performance and costs with fluctuating workloads. This correlation further makes memory configuration a non-trivial task. On that account, an input-aware function memory allocator not only improves developer productivity by completely hiding resource-related decisions but also drives an opportunity to reduce resource wastage and offer a finer-grained cost-optimised pricing scheme. Therefore, we present MemFigLess, a serverless solution that estimates the memory requirement of a serverless function with input-awareness. The framework executes function profiling in an offline stage and trains a multi-output Random Forest Regression model on the collected metrics to invoke input-aware optimal configurations. We evaluate our work with the state-of-the-art approaches on AWS Lambda service to find that MemFigLess is able to capture the input-aware resource relationships and allocate upto $82$\% less resources and save up to $87$\% run-time costs.
\end{abstract}

\begin{IEEEkeywords}
Serverless Computing, Function-as-a-Service, function configuration, input-awareness, constraint optimisation
\end{IEEEkeywords}

\section{Introduction}
The serverless computing paradigm is the latest cloud-native development model that enables application execution without the management of underlying resources. \textit{Serverless} promotes the idea that a developer should be less concerned about the servers or infrastructure and focus more on productivity that adds value to the business. This shift of responsibility means offloading resource management tasks to the cloud service provider (CSP), such as resource allocation, application scaling and software updates. In the serverless landscape \cite{CNCF}, Function-as-a-Service (FaaS) emerged as a microservices-inspired, event-driven execution model where \textit{function(s)} are integrated with additional Backend-as-a-Service (BaaS) offerings like storage, networking and database services, to set-up an application. A serverless \textit{function} is a stateless code fragment, executed on-demand within lightweight virtual machines (VM), microVMs or containers for short-term duration, and bills its resources as per usage.

In 2014, Amazon Web Services (AWS) introduced AWS Lambda \cite{BerkleyView2019}, \cite{Lambda} as its first FaaS offering, and since then, a range of FaaS services have emerged, including Google Cloud Functions \cite{GCF}, Azure Functions \cite{Azure}, and many open-source implementations such as OpenFaaS \cite{OpenFaaS}, Knative \cite{Knative} and OpenWhisk \cite{OpenWhisk}. In addition to serverless attributes such as on-demand scalability, zero idle-resource costs, and no resource management, FaaS uniquely features scale-to-zero capability where function resources are released after an extended period of inactivity, endorsing a multi-tenant resource-sharing and pay-per-use pricing model. FaaS has increasingly found its relevance in a variety of use cases like 
video streaming platform \cite{ServerlessVideo}, multi-media processing \cite{NetflixAWS}, CI/CD pipeline \cite{CapitalOne}, AI/ML inference task \cite{AI-ML-AWS}, and Large-Language-Model (LLM) query processing \cite{LLM-Medium-AWS}.

The operational model of FaaS hides the complex infrastructure management from end users and does not signify the absence of servers. A serverless function still requires resources, including computing, network and memory, for a successful execution. In the current FaaS implementations, a developer is responsible for requesting the right combination of resources to guarantee successful function execution. However, service providers only expose a small set of resource knobs, usually memory 
\footnote{We refer to FaaS platforms like AWS Lambda that allow developers to provide only memory configuration and allocate CPU, network bandwidth, etc., in a proportional fashion.} 
with proportionally allocated CPU, disk I/O, network bandwidth, etc. \cite{Sizeless}. Prior studies \cite{PredictingTheCostsOfWorkflows}\cite{StepConf}\cite{WithGreatFreedomComesGreatOpportunity} have identified that a higher memory configuration speeds up function execution and has a significant impact on its start-up performance and costs. However, the execution speedup is non-linear and has a diminishing marginal improvement with increasing memory allocations \cite{COSE.V2}. With limited observability into short-running functions and unaware of function performance, developers usually resort to speculative decisions for memory configuration or make experience-based ad-hoc decisions with an expectation to fulfil service level objectives (SLO) \cite{CPUTAMS}. To validate such developer behaviour, an industry insight \cite{Datadog2020} reports the ease of controlling function execution duration via memory configuration, while 47\% of production-level functions still run with the default memory configuration without exploring the entire configuration space. Additionally, selecting an optimal memory configuration from an exponentially large search space requires a careful understanding of the correlation between function performance and resource requirements. Hence, configuring the function with the right amount of memory that guarantees shorter execution times and lower execution costs is an intricate task. 

\begin{figure*}[t]
    \centering
       
    \begin{subfigure}{0.28\textwidth}
        \includegraphics[width=\textwidth]{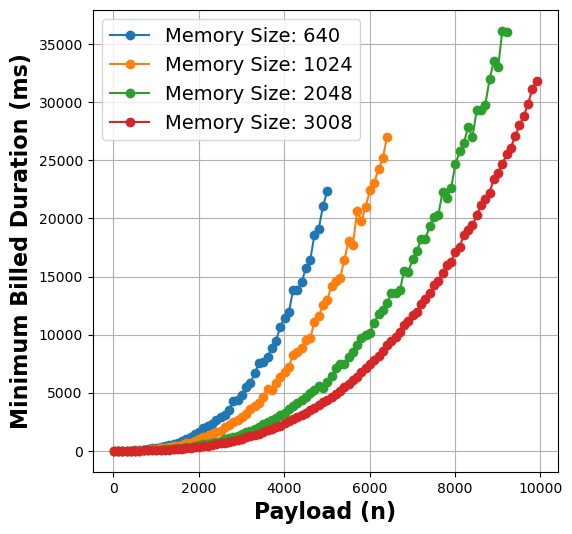}
        \caption{Payload vs Duration \textit{matmul} function metrics}
        \label{fig:matmul_payload_duration}
    \end{subfigure}
    \hfill
    \begin{subfigure}{0.28\textwidth}
        \includegraphics[width=\textwidth]{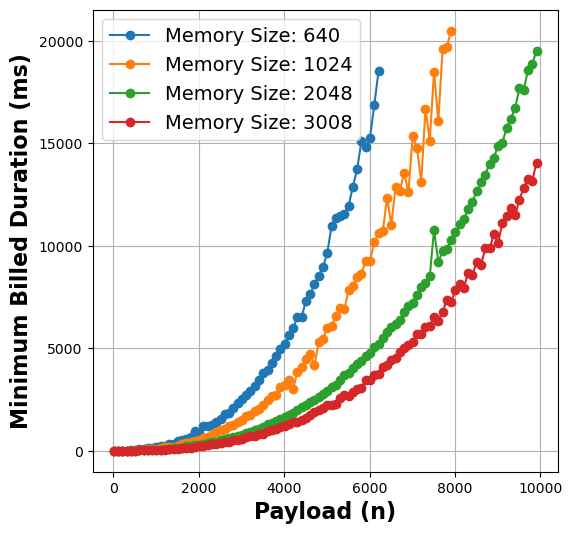}
        \caption{Payload vs Duration \textit{linpack} function metrics}
        \label{fig:linpack_payload_duration}
    \end{subfigure}
    \hfill
        \begin{subfigure}{0.28\textwidth}
        \includegraphics[width=\textwidth]{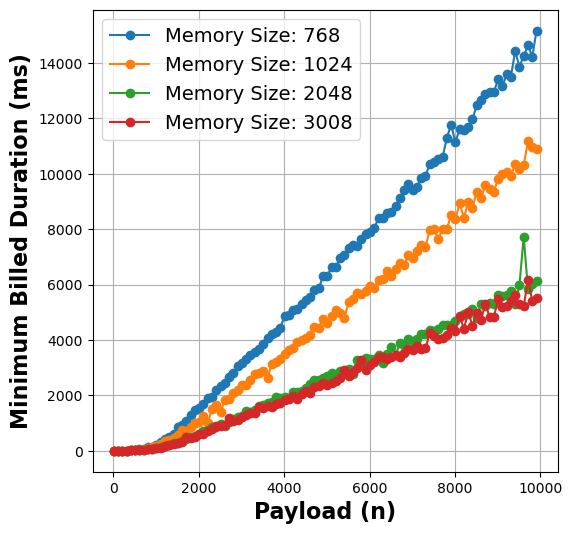}
        \caption{Payload vs Memory Utilisation \textit{graph-mst} function metrics}
        \label{fig:mst_payload_duration}
    \end{subfigure}
    \vskip\baselineskip
    \begin{subfigure}{0.28\textwidth}
        \includegraphics[width=\textwidth]{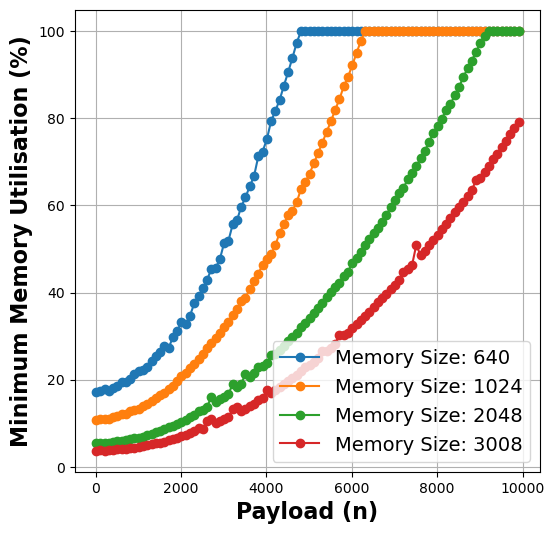}
        \caption{Payload vs Memory Utilisation \textit{matmul} function metrics}
        \label{fig:matmul_payload_memory}
    \end{subfigure}
    \hfill
    \begin{subfigure}{0.28\textwidth}
        \includegraphics[width=\textwidth]{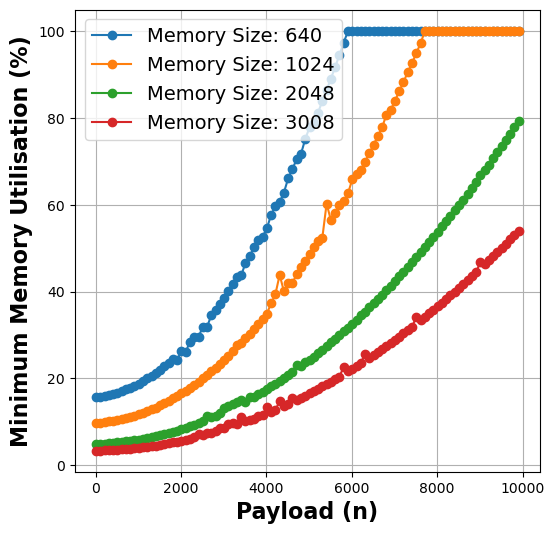}
        \caption{Payload vs Memory Utilisation \textit{linpack} function metrics}
        \label{fig:linpack_payload_memory}
    \end{subfigure}
    \hfill
        \begin{subfigure}{0.28\textwidth}
        \includegraphics[width=\textwidth]{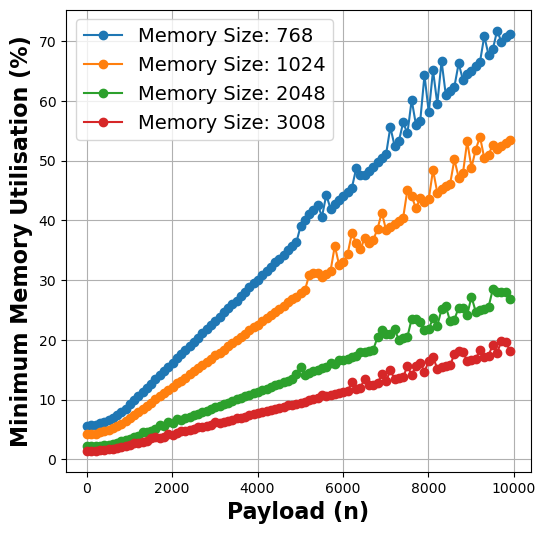}
        \caption{Payload vs Memory Utilisation  \textit{graph-mst} function metrics}
        \label{fig:mst_payload_memory}
    \end{subfigure}
    \caption{Function Metrics Insight - Payload vs Memory Utilisation vs Billed Duration}
    \label{fig:metrics_insight}
\end{figure*}

\begin{table}
\centering
\caption{List of collected function metrics}
\label{tab:function_metrics}
\resizebox{\columnwidth}{!}{%
\begin{tabular}{|p{0.35\linewidth} | p{0.6\linewidth}|}
\hline
\textbf{Metric Name} & \textbf{Description} \\ \hline
request\_id & unique function invocation ID \\ \hline
payload & function input parameter(s) \\ \hline
memory\_size & amount of memory allocated to function \\ \hline
memory\_utilisation & maximum memory measured as a percentage of the memory allocated to the function \\ \hline
memory\_used & measured memory of the function sandbox \\ \hline
billed\_duration & function execution time rounded to nearest millisecond \\ \hline
billed\_mb\_ms & total billed Gb-s, a pricing unit for function \\ \hline
cold\_start & function cold start (true/false) \\ \hline
init\_duration & amount of time spent in the init phase of the execution environment lifecycle \\ \hline
function\_error & any function run-time error \\ \hline
\end{tabular}%
}
\end{table}

Recent research \cite{WithGreatFreedomComesGreatOpportunity}\cite{OFC}\cite{Cypress}\cite{ParrotFish} that optimise the function resource allocation process has highlighted a drastic impact of input parameters on its performance. Additionally, a static memory configuration is used for concurrent function invocations while expecting similar performance for distinct function inputs. Therefore, setting a static memory configuration for all function invocations, regardless of their input, leads to a fluctuating performance with varying workload and input arguments. This performance unpredictability demands an input-argument-aware approach in determining the memory configuration for function invocations that balances execution cost and running time while reducing excess resource allocation. This input-based memory configuration has a two-fold effect of providing a more autonomous developer experience and a chance for CSPs to maximise resource utilisation and deliver a finer-grained, cost-effective pricing model for users. Additionally, existing efforts \cite{WithGreatFreedomComesGreatOpportunity}\cite{OFC}\cite{Cypress}\cite{ParrotFish} to configure function resources either focus on an average-case function execution to recommend maximum used memory/resources or propose to re-run their solution for specific input parameters to optimise the memory allocation process. This may lead to higher run-time costs and resource wastage and on the other hand, running multiple models for previously unseen input values extends the data collection process as well as increases the model training and tuning complexity. Therefore, a solution is warranted that captures the relationship of input parameters with function resources to precisely model and predict the required memory configuration for successful execution and reducing excess resource allocation.

To this end, we present MemFigLess, an end-to-end estimation and function memory allocation framework that makes input-aware memory configuration decisions for a serverless function. MemFigLess takes as an input the function details, such as the representative function input arguments, expected running time and cost SLOs and a range of memory allocations to explore. The framework executes an offline profiling loop to take advantage of a robust tree-based ensemble learning technique, multi-output Random Forest Regression (RFR), which analyses the relationship between input parameters and other function metrics such as execution time, billed cost, and function memory requirement. The RFR model is then exploited in an online fashion to make an optimal selection of memory configuration for individual function invocations. Additionally, the framework provides a feedback loop to re-train the model in a sliding-window manner with a new set of collected metrics to capture the performance variation. 

\section{Motivation}
\label{section: motivation}

To identify and establish the effect of input parameters on a function's memory requirement and execution time, 
we experiment with the industry-leading FaaS platform, AWS Lambda \cite{Lambda} and conduct a large-scale initial study by deploying three benchmark functions, \begin{enumerate*}
    \item matrix multiplication (\textit{matmul}), 
    \item graph minimum spanning tree (\textit{graph-mst}), and
    \item \textit{linpack} 
\end{enumerate*} from \cite{FuncitonBench}. A CPU-bound function, \textit{matmul}, calculates the multiplication of a $n * n$ matrix, whereas \textit{linpack} measures the system's floating-point computing power by solving a dense $n$ by $n$ system of linear equations. \textit{graph-mst} is a scientific computation offloaded to serverless functions that
generates a random input graph of $n$ nodes using Barabási–Albert preferential attachment and processes it with the minimum spanning tree algorithm. To observe the effect of payload (i.e., input parameters) on the performance of benchmark functions, we execute them with input set $N = \{ n | 10\leq n \leq 10000, n \leftarrow n + 100\}$ and vary the memory configuration of the function $M = \{m | 128 \leq m \leq 3008, m \leftarrow m + 128 \}$ MB. We take advantage of Amazon CloudWatch \cite{CloudWatch} as a monitoring solution to gather function-level performance metrics.

We collect relevant function metrics as described in Table \ref{tab:function_metrics} and plot the function payload against the minimum billed duration and minimum memory utilised in Fig. \ref{fig:metrics_insight}. 

\textbf{Insight 1:} \textit{There is a \textbf{strong} correlation between the function payload and execution time that varies in proportion to distinct memory allocations.}

We find a strong correlation between the function payload and the corresponding billed duration for all the benchmark functions. It can be inferred from Fig. \ref{fig:matmul_payload_duration} - \ref{fig:mst_payload_duration} that the minimum billed duration, i.e., the execution time of a function, is directly proportional to its input and has a tendency to increase with increasing payloads at distinct memory configurations. Therefore, we can infer that different memory configurations lead to proportional resource allocations and thus, the function performance, i.e., execution time, also varies in proportion to these available resources. This complex relation of payload-dependent execution time further aggravates the overall function run-time cost as it is calculated based on the execution time and allocated memory.

\textbf{Insight 2:} \textit{There is a \textbf{positive} correlation between the payload and minimum memory utilised for successful execution of the function, which has a direct and complex relationship with resource wastage and run-time cost.}

In Fig. \ref{fig:matmul_payload_memory} - \ref{fig:mst_payload_memory}, we observe the effect of function payload on memory utilisation. A higher memory utilisation is observed for all the payload values at the lower memory allocations, while a lower memory utilisation can be seen at higher memory allocations for all benchmark functions. Therefore, an under-utilised function resource depicts an inherent resource wastage, where the function payload has a direct effect on it. However, the relationship may not be the same for a function and payload combination, and thus requires thoughtful resource allocation to reduce excess resource wastage and associated run-time costs. In addition to this, it is well established in previous research studies \cite{COSE.V2}\cite{ParrotFish}\cite{Lachesis} that a function experiences a performance speed up with additional resources and hence, a complex association exists between the function resource allocation and pricing schemes. Therefore, this work attempts to address the key challenges identified in the motivation study.

\section{Related Work}
\label{section:related_work}


The authors in \cite{Sizeless} propose a multi-regression model generated on synthetic function performance data and use it to predict execution time and estimate cost at distinct memory configurations. Similarly, \cite{MAFF} explores search algorithms like linear/binary search and gradient descent to determine the optimal configuration for cost-focused or balanced optimisation goals. However, none of them considers the effect of payload on function performance and resource demands. Additionally, they either rely on synthetic data or perform repeated search across configurations. 

Jarachanthan et al. \cite{Astrea} explore the performance and cost trade-off of the function at different configurations for Map-Reduce-style applications. They propose a graph-theory-based job deployment strategy for Directed Acyclic Graph (DAG) based applications to optimize the resource configuration parameters. Wen et al. \cite{StepConf} focus on multicore-friendly programming for workflows to estimate the inter- / intra-function parallelism based on weighted sub-SLOs. Safaryan et al. \cite{SLAM} focus on the SLO-aware configuration of workflows and use a max-heap data structure to find a configuration repeatedly. However, these works address function workflows to either fit a specific application style or configure function parallelism and fall short of considering the payload effect.

The work in \cite{TimeCostMemoryConfig} introduces an urgency-based heuristic method where a particle swarm optimization technique is used for time/cost trade-off. Lin and Khazaei \cite{ModellingandOptimisationOfPerformanceCost} propose a probability-refined critical path greedy algorithm for selecting the optimal function configuration from the profiled data. The authors in \cite{CPUTAMS} discuss the concept of CPU time sharing and resource scaling with memory configuration to propose a memory-to-vCPU Regression model. However, these works model the behavior of functions and do not consider the payload effect on function performance or assume an homogeneous function resource relationships. Raza et al. \cite{COSE.V2} explore a Bayesian Optimisation (BO) based model to optimally configure and place functions considering their execution time and cost. However, they disregard the payload effect on resource configuration and performance.


In \cite{PredictingTheCostsOfWorkflows}, the authors establish a relationship between input/output parameters and response time to predict their distribution via offline Mixture Density Networks (MDN) and utilise online Monte-Carlo simulations to estimate best workflow costs. However, these estimates consider fixed memory configuration for performance and cost prediction. Researchers in \cite{OFC} discuss a high correlation between input arguments and memory configuration for Extract-Transform-Load (ETL) pipeline functions using Decision Trees (DTs) and re-claim unused memory to resize worker-node cache for locality-aware function executions. However, this is application specific and focuses on cache memory management. Bhasi et al. \cite{Cypress} present an input-size sensitive resource manager that performs request batching, request re-ordering and rescheduling to minimise resource consumption and maintain a high degree of SLO. However, they focus on request management and discard the opportunity of input-sensitive function configuration.

Bilal et al. \cite{WithGreatFreedomComesGreatOpportunity} re-visits Bayesian Optimisation models and Pareto front prediction, among few to discuss a complex challenge of input dependency for resource allocation. However, they primarily answer design space questions to showcase potential opportunities for flexible resource configuration. Sinha et al. \cite{Lachesis} presents an online supervised learning model to allocate the minimum amount of CPU resources for individual invocations based on input characteristics and function semantics. However, they introduce an overhead of supervising individual invocations for SLOs in addition to adjusting only CPU resources.

Moghimi et al. \cite{ParrotFish} criticise available function right-sizing tools for reducing developer efficiency and explores a characterisation-driven modelling tool that takes advantage of parameter fitting, adaptive sampling and execution logs to find the right function configuration. Although they consider relative payloads while suggesting optimal configuration, they recommend re-execution of their model for individual payloads. This introduces a run-time overhead to find payload specific configuration in terms of profiling time and costs. Therefore, with this work we attempt to address payload-aware function memory configuration, where other resources are proportionally allocated, to satisfy run-time performance constraints and make these configuration decisions online.

\section{Problem Formulation and Architecture}
\label{sec:system_arch_prob_formulation}
In this section we formally describe the function configuration problem and introduce the system architecture.

\subsection{Problem Formulation}

In current FaaS environments, developers must configure memory settings (referring to FaaS platforms like AWS Lambda)
for their functions to ensure successful execution. However, determining the appropriate memory configuration is challenging due to factors like variability in function input or payload characteristics (e.g., input size, type, and number of inputs) and invocation frequency, which significantly impact resource demands, run-time performance and cost. To handle this, developers generally make ad-hoc decisions to either configure platform defaults \cite{Datadog2020} or speculate the right-sizing of functions based on past experience. These decisions lead to sub-optimal resource allocations, over- or under-provisioning, which may result in resource wastage, added run-time costs, and throttled function performance. Additionally, FaaS platforms scale a function with static resource configuration with fluctuating workload which further makes the resource scaling and allocation challenging. 

Existing research \cite{Sizeless} \cite{COSE.V2}\cite{AWSPowerTuning}\cite{Lachesis} have repeatedly accentuated the complex relationship of function resource demand, execution time guarantee and run-time cost, which is further complicated when considering function payload \cite{ParrotFish}. Therefore, we formulate the problem of payload-aware function configuration as a multi-objective optimisation (MOO) problem where the objective is to select a memory configuration that guarantees a successful execution within an advertised function deadline, while reducing excess resource allocation and run-time costs for incoming function invocations with varying payloads. The problem can be mathematically represented as Eq. \ref{eq:objective}, such that the constraints Eq. \ref{eq:constraints} are satisfied. 

\begin{equation}
\label{eq:objective}
    \min_{m \in M} G(m,P) = (C_f(m, P),T_f(m, P))
\end{equation}

\textbf{Subject to:}

\begin{align}
\label{eq:constraints}
& T_f(m, P) \leq D, \\
& C_f(m, P) = T_f(m,P)*C_m + \beta \leq B, \\
& \text{Success}(m, P) = 1.
\end{align}

In FaaS, a function $f$ may expect a number of inputs, $P = \{P_1,\dots,P_n\}$ where an input belongs to a range of values $P_n = \{p^{min}_n, \dots, p^{max}_n\}$, either continuous or discrete, which influences the function memory requirements $m \in M = \{m^{min},\dots,m^{max}\}$. We define the objective of payload-aware memory configuration to minimise the run-time cost $C(m, P)$ and the function execution time $T(m, P)$ at memory allocation $m$ and payload(s) $P$, such that a function executes successfully within the specified deadline $D$ and budget $B$ constraints. The run-time cost $C_f(m,P)$ is directly proportional to the execution time $T_f(m,P)$ and the cost of memory configuration $C_m$ plus a constant invocation amount, $\beta$ (provider dependent).

\begin{figure}[t]
    \centering
    \begin{subfigure}{0.35\textwidth}
        \includegraphics[width=\textwidth]{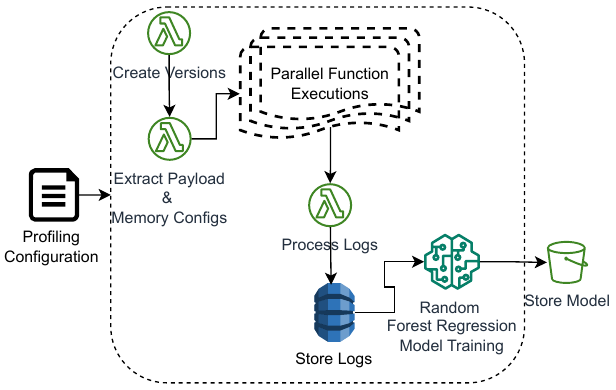}
    \caption{MemFigLess offline profiling and data collection workflow.}
        \label{fig:offline_stage}
    \end{subfigure}
    \hfill
        \begin{subfigure}{0.35\textwidth}
        \includegraphics[width=\textwidth]{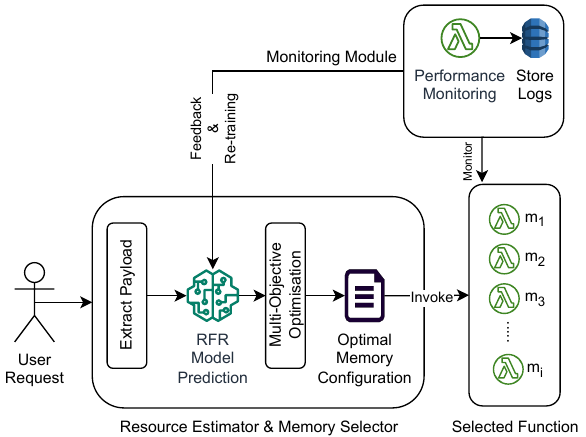}
        \caption{MemFigLess Online workflow.}
        \label{fig:online_stage}
    \end{subfigure}
    
    \label{fig:memfigless_architecture}
    \caption{MemFigLess System Architecture}
\end{figure}

\subsection{System Architecture}
The proposed MemFigLess system architecture consists of both offline and online components designed to optimize memory allocation for FaaS offerings based on the payload. The model of the proposed framework is a $MAPE$ control loop, i.e., \textit{Monitor}, \textit{Analyse}, \textit{Plan} and \textit{Execute}. In the online stage, a periodic monitoring of function performance metrics is done. It is then analysed by the resource manager via a feedback loop to plan and execute the resource allocations that meet the performance SLOs.

\subsubsection{Offline Profiling and Training Module}
\label{subsec:offline_profiling}
This module is responsible for profiling the functions and training the Random Forest Regressor (RFR) model, Fig. \ref{fig:offline_stage}. Functions are executed with a variety of inputs to collect data on their performance and resource usage. Metrics such as input size, number of inputs, memory consumption, execution time and billed execution unit are recorded. The collected metrics are stored in a structured format to serve as training data for the model. A tree-based ensemble learning RFR model is trained on the collected data to learn the relationship between the function’s input/payload and its memory requirement and execution time. The model captures the impact of input size distribution on resource usage and is used for online payload-aware memory estimation. 

\subsubsection{Online Prediction and Optimization Module}
\label{subsec:online_prediction}
This module leverages the trained RFR model to select the optimal function memory based on the model prediction and constraint optimisation, and invoke functions in real-time, as shown in Fig. \ref{fig:online_stage}. Incoming function requests are analyzed to extract payloads which are fed to the trained RFR model to predict the execution time at distinct function memory configurations. The selected memory configurations, based on SLO constraints, are used for online constraint optimisation, either cost or execution time, provided by the user. This helps in reducing the potential resource wastage and overall cost of execution.

\subsubsection{Dynamic Resource Manager}
\label{subsec:resource_manager}

This component is embedded in the online prediction module to handle the invocation and allocation of resources based on the predictions made by the RFR model. The resource manager dynamically allocates function resources or selects the available function instance based on the constraint-optimised memory selection, ensuring minimal wastage and optimal performance. A continuous monitoring and feedback mechanism is also incorporated to improve the accuracy and performance of the system over time. This module can monitor and log the performance of functions during execution within a configured $monitoring\_window$ to ensure that the model captures performance fluctuations periodically. The gathered performance data is leveraged by the training module to periodically re-train and improve the RFR model predictions.

\section{Multi-Output Random Forest Regression}
\label{sec:rfr}
To accurately select the memory configuration of serverless functions, a Random Forest (RF)-based regression algorithm can be employed. The Random Forest, initially presented by Breiman \cite{RandomForest}, is one of the most popular supervised machine learning (ML) algorithms and has been successfully applied to both classification and regression in many different tasks, such as virtual machine (VM) resource estimation \cite{RFCloudCPU}, VM resource auto-scaling \cite{RFAutoscaler23} and computer vision applications \cite{RFCompVision}. The RF algorithm uses a combination of DTs to model complex interactions between input parameters and identify patterns in the data. It works by training multiple DTs on subsets of input parameters and then aggregating their predictions to generate the final estimation. This method has been demonstrated to have the ability to accurately approximate the variables with nonlinear relationships and also have high robustness performance against outliers. In addition, compared to other ML techniques, e.g., Artificial Neural Networks (ANN), Support Vector Machine (SVM), Deep Learning and Reinforcement Learning, it only needs a few tunable parameters and therefore requires low effort for offline model tuning. Furthermore, the algorithm can handle noisy or incomplete input data, and reduces the risk of overfitting which might occur with other ML algorithms \cite{AIBlock_Golec}.

The RF model is an ensemble-learning method that can be modeled as a collection of DTs. A DT makes a prediction for the input feature vector $\Vec{x} \in F$, where $F$ represents a subset of $\kappa$-dimensional feature space \cite{RandomForest}. A DT recursively partitions the feature space $F$ into $L$ terminal nodes or leaves that represent the region $R_l, 1 \leq l \leq L$ where every possible feature vector $\Vec{x}$ may belong. Therefore, an estimation function $f(\Vec{x})$ for a DT can be summarised as Eq. \ref{eq:prediction}, where $I(x, R_l)$ represents the indicator function of whether the feature vector $\Vec{x}$ belongs to region $R_l$. The function $f(\Vec{x})$ indicates how the DTs return the value of leaf corresponding to the input $\Vec{x}$ and typically learns a response variable $c_l$ for each region $R_l$ where $\Vec{x}$ belongs, to assign an average value to that region in the regression tree \cite{RFAutoscaler23}.

\begin{align} 
\label{eq:prediction}
f(\mathbf {\overrightarrow{x}})=&\sum _{l=1}^{L}c_{l} I(\mathbf {\overrightarrow{x}},R_{l})
\end{align} 
\begin{align} 
\label{eq:indicator}
I(\mathbf {\overrightarrow{x}},R_{l})=&\begin{cases} 1; & \mathbf {\overrightarrow{x}}\in R_{l}\\ 
0; & \mathbf {\overrightarrow{x}} \notin R_{l} \end{cases}
\end{align} 

However, a more interpretative representation is Eq. \ref{eq:fullDT} where $c^{full}$ is the average of all learned response variables during the training and $C(\Vec{x}, k)$ is the contribution of the $k^{th}, 1 \leq k \leq \kappa$ feature in $x$.

\begin{align} 
\label{eq:fullDT}
f(\mathbf {\overrightarrow{x}})=c^{full}+\sum _{k=1}^{K} C(\mathbf {\overrightarrow{x}},k)
\end{align}

Therefore, the average prediction $F(\Vec{x})$ for a RFR model over an ensemble of DTs can be summarised as Eq. \ref{eq:randomforest}, where $S$ is the number of DTs, $C_s^{full}$ is the contribution of $k^{th}$ feature in vector $x$ in $j$-th DT.

\begin{align} 
\label{eq:randomforest}
F(\overrightarrow{x})=\frac {1}{S} \sum _{s=1}^{S} c_{s}^{full}+\sum _{k=1}^{\kappa} \left ({\frac {1}{S}\sum _{s=1}^{S} C_{s}(\mathbf {\overrightarrow{x}},k) }\right)
\end{align} 

To apply the RFR in a serverless framework, first, we need to collect relevant function performance metrics at distinct representative payloads. This data is gathered through a series of experiments, Sec. \ref{subsec:offline_profiling}, where each input parameter is varied, and the resulting function metrics like memory consumption are measured. Once the performance data is gathered, we train a random forest regression model, as outlined in Sec. \ref{subsec:offline_profiling}. In our problem context, the input vector $\Vec{x}$ has multiple components, which are \textit{total\_memory} allocation $m$ and function \textit{payload(s)}, $P$.  This RFR model predicts the \textit{billed\_duration} and \textit{memory\_utilisation}, which directly computes run-time cost and execution duration, and \textit{function\_error\_status} for determining successful execution. These predictions align with the objectives of function run-time cost $C_f(m, P)$ and execution time $T_f(m, P)$ while ensuring a successful execution.

To address the conflicting objectives i.e., executing a function within a deadline, $D$ and with a run-time budget, $B$, commonly, the concept of Pareto dominance and Pareto optimality are used \cite{ParetoMOO}. This optimisation is integrated with online prediction module \ref{subsec:online_prediction}, to select a payload-aware and constraint-optimised memory configuration. Pareto dominance is a method for comparing and ranking the decision vectors. A vector $\Vec{x_u}$ is said to dominate vector $\Vec{x_v}$ in the Pareto sense, if an objective vector $G(\Vec{x_u})$ is better than $G(\Vec{x_v})$ across all objectives, with atleast one objective where $G(\Vec{x_u}) > G(\Vec{x_v})$, strictly. A solution $\Vec{\hat{x}}$ is said to be Pareto optimal if there does not exist any other solution that dominates it and then the objective $G(\Vec{\hat{x}})$ is known as Pareto dominant vector. Therefore, a set of all Pareto optimal solutions is called Pareto set and corresponding objective vectors are said to be on Pareto front.

We approach our MOO by transforming the multi-objective problem into a single objective by employing the classical Weighted Aggregation Method (WAM), where a function operator is applied to the objective vector $G(\Vec{x})$. As a user is responsible for providing the relative importance of objectives, we select a linear weighted combination as the utility function for objective optimisation. Therefore, the final optimisation problem can be simplified as Eq. \ref{eq:MOO} where $J_z$ represents $z^{th}$ objective with a relative weight of $w_z$ and a weighted combination of all the objectives are jointly minimised.

\begin{align}
\label{eq:MOO}
    \min_{\overrightarrow{x}} Z = \sum \limits_1^z w_z*J_z(\overrightarrow{x})
\end{align}

\textbf{Subject to:}

\begin{align}
    w_z \geq 0 \ {\rm ;} \ \sum \limits_1^z w_z = 1
\end{align}

Once the Pareto front is obtained using the discussed MOO, we select the memory configuration that is cheapest in terms of resource allocation, i.e., the lowest memory configuration from the Pareto optimal solutions and execute it via the dynamic resource manager, Sec. \ref{subsec:resource_manager}.

\section{Performance Evaluation}
\label{sec:evaluation}

In this section, we briefly discuss the implementation along with experimental setup, model parameters, and perform an analysis of the proposed RFR-based framework compared to other complementary solutions.

\subsection{Implementation and System Setup}
\label{subsec:system_setup}

We setup our proposed solution using AWS serverless services \cite{AWS_Serverless_Suite}, such as AWS Lambda, AWS DynamoDB, AWS Step Functions Workflow and Amazon Simple Storage Service (S3) for an end-to-end serverless function configuration solution. The framework can be assumed a CSP service where users can subscribe to it for an end-to-end optimisation, based on the desired deadline and run-time cost of the individual function. The offline step is implemented as a Step Functions Workflow that takes function details such as resource name, memory configurations to explore, number of profiling iterations and the representative payload(s) as a $json$ input file. This information is used to create and execute a function with different payloads at distinct configurations, and collect the performance data using AWS CloudWatch \cite{CloudWatch}. However, the payload for different functions may be of different types and thus, the workflow utilises the AWS S3 service to fetch any stored representative payload. The individual workflow tasks of creating, executing and updating the function in addition to log collection and processing, are implemented as AWS Lambda functions. All the functions are configured with a $15$ minutes timeout, $3008$ MB (maximum free tier) memory configuration, and $512$ MB ephemeral storage, to avoid any run-time resource scarcity. The processed logs are stored in AWS DynamoDB, a persistent key-value datastore. These logs are utilised by RFR training function and the trained model is placed in S3 storage for online estimation.

The online step makes use of the trained RFR model which is also implemented as a function, assuming shorter executing functions. The RFR model is implemented using Scikit-Learn \cite{scikit-learn}, a popular ML module in Python. This function loads the RFR model for inference, estimates resources, performs the optimisation and invokes the selected function configuration. In addition, performance monitoring can be scheduled to collect and store the logs in the key-value datastore and a \textit{monitoring\_window} can be setup for periodic log processing and model re-training.

\begin{figure*}[t]
    \centering
    \begin{subfigure}{0.22\textwidth}
        \includegraphics[width=\textwidth]{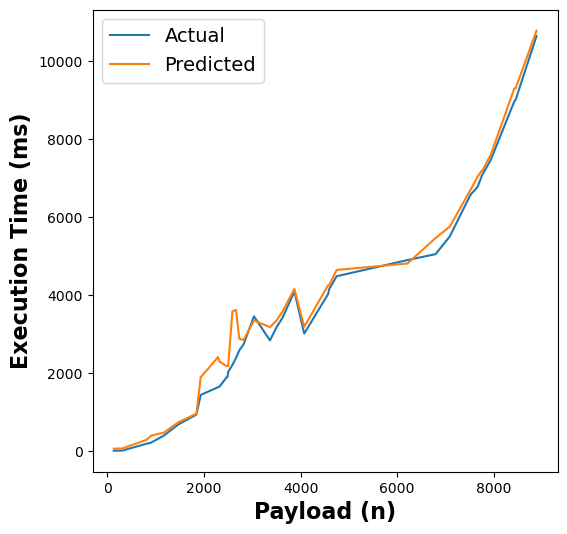}
        \caption{RFR Execution time estimates for \textit{linpack}}
        \label{fig:linpack_rfr}
    \end{subfigure}
    \hfill
    \begin{subfigure}{0.22\textwidth}
        \includegraphics[width=\textwidth]{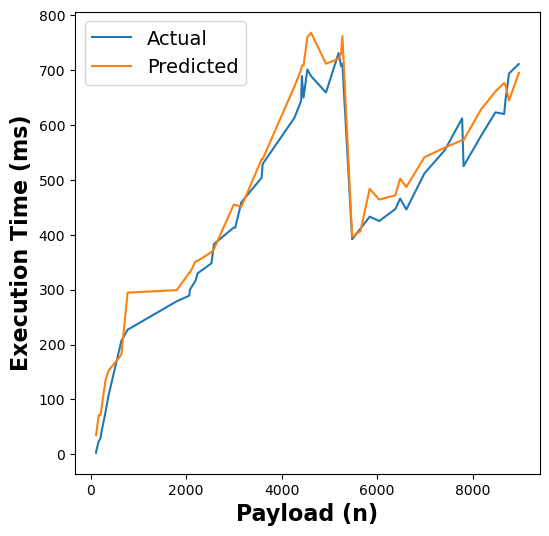}
        \caption{RFR Execution time estimates for \textit{graph-pagerank}}
        \label{fig:pagerank_rfr}
    \end{subfigure}
    \hfill
        \begin{subfigure}{0.22\textwidth}
        \includegraphics[width=\textwidth]{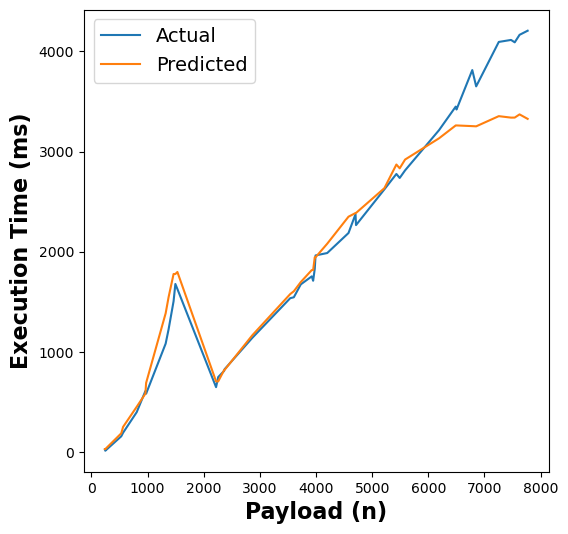}
        \caption{RFR Execution time estimates for \textit{graph-bfs}}
        \label{fig:bfs_rfr}
    \end{subfigure}
    \hfill
    \begin{subfigure}{0.22\textwidth}
        \includegraphics[width=\textwidth]{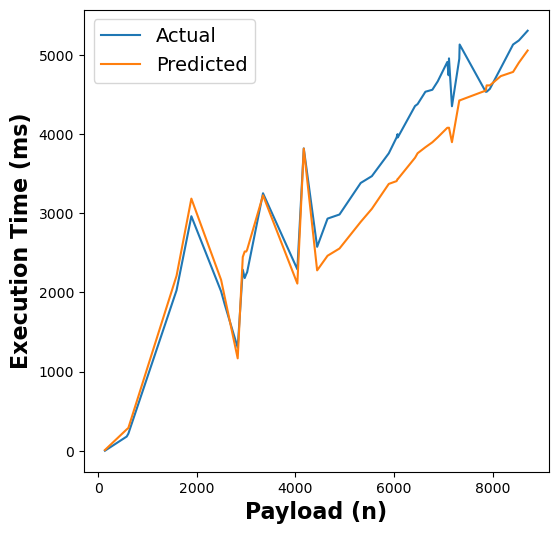}
        \caption{RFR Execution time estimates for \textit{graph-mst}}
        \label{fig:mst_rfr}
    \end{subfigure}
    \vskip\baselineskip
    \begin{subfigure}{0.22\textwidth}
        \includegraphics[width=\textwidth]{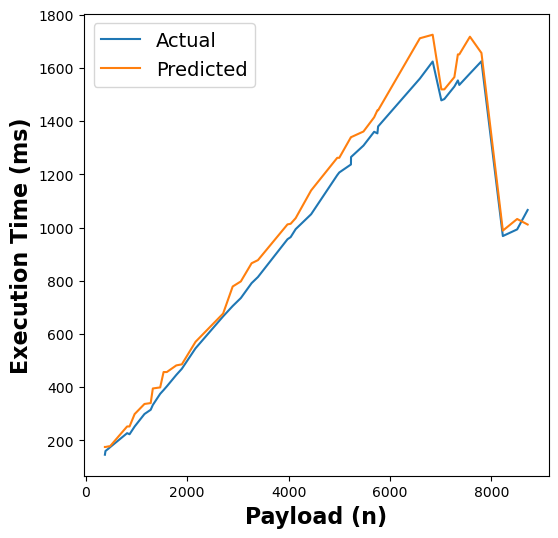}
        \caption{RFR Execution time estimates for \textit{chameleon}}
        \label{fig:chameleon_rfr}
    \end{subfigure}
    \hfill
        \begin{subfigure}{0.22\textwidth}
        \includegraphics[width=\textwidth]{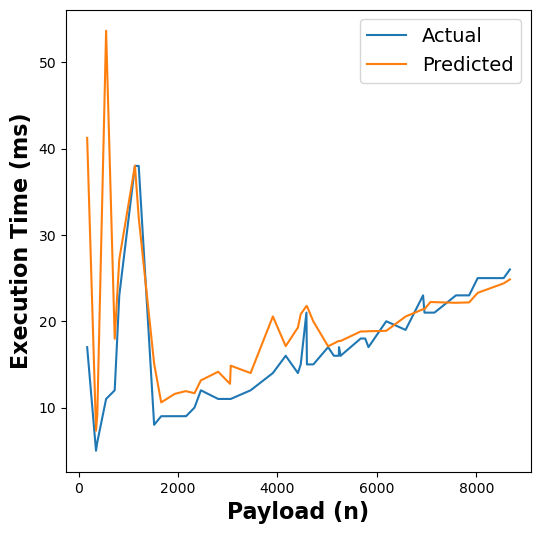}
        \caption{RFR Execution time estimates for \textit{dynamic-html}}
        \label{fig:html_rfr}
    \end{subfigure}
        \hfill
    \begin{subfigure}{0.22\textwidth}
        \includegraphics[width=\textwidth]{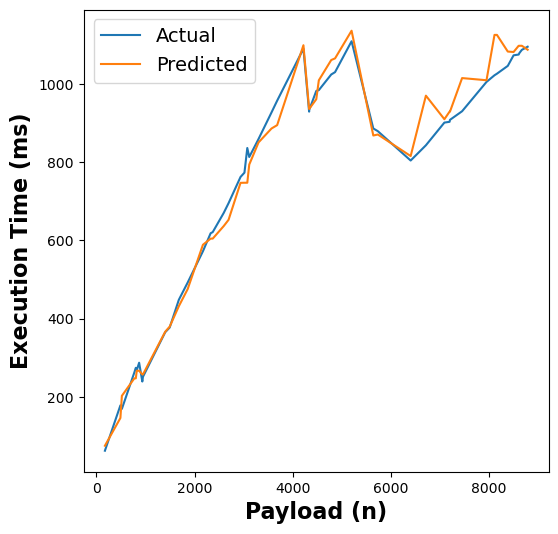}
        \caption{RFR Execution time estimates for \textit{pyaes}}
        \label{fig:pyaes_rfr}
    \end{subfigure}
    \hfill
    \begin{subfigure}{0.22\textwidth}
        \includegraphics[width=\textwidth]{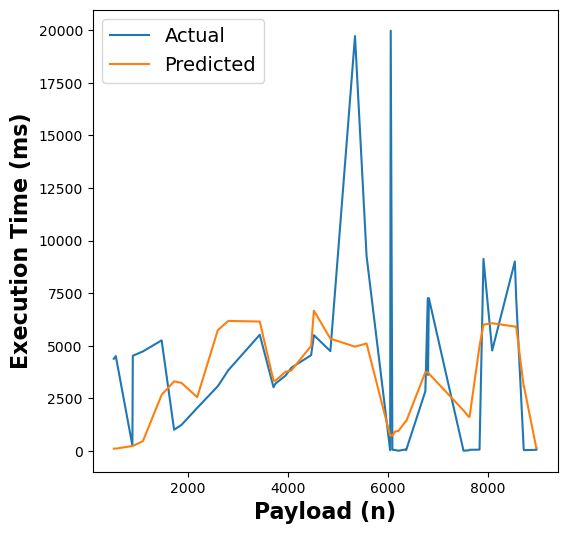}
        \caption{RFR Execution time estimates for \textit{matmul}}
        \label{fig:matmul_rfr}
    \end{subfigure}
    
    \caption{RFR Model Payload-Aware Execution Time Prediction.}
    \label{fig:rfr_prediction}
\end{figure*}

The RFR model assumes that the payloads provided in the offline step are representative of actual payload and therefore, inference at any anomalous/outlier value is defaulted to $3008$ MB or the estimated configuration for the smallest payload value seen. The inference model expects a function deadline $D$, run-time budget $B$ and their relative importance, $w_z$, for optimisation and configuration selection. Based on the initial analysis we configure the deadline $D$ as the mean function execution time, and the run-time budget $B$ as the mean execution cost across memory and payload combination. 

We perform our experimental analysis on a range of functions implemented in Python v3.12, taken from serverless benchmarks \cite{FuncitonBench} and \cite{SebsBenchmark}, including CPU/memory intensive (\textit{matmul, linpack, pyaes}), scientific functions (\textit{graph-mst, graph-bfs, graph-pagerank}) and dynamic website generation (\textit{chameleon, dynamic-html}). The explored functions and required payloads are listed in Table \ref{tab:function_payload} and the experimental payload values range between $[10, 10000]$ with a step\_size of $200$, for individual variables. This \textit{step\_size} was randomly chosen for the experiments and must be provided by the user for the granularity of experiments and model generation.

\begin{table}
\caption{List of functions and payload value}
\label{tab:function_payload}
\resizebox{\columnwidth}{!}{%
\begin{tabular}{|p{0.35\linewidth} | p{0.6\linewidth}|}
\hline
\textbf{Function Name} & \textbf{Payload} \\ \hline
matmul & $n$, size of matrix \\ \hline
linpack & $n$, number of linear equations to solve \\ \hline
pyaes & $\{n, m\}$, length of message to encrypt and number of iterations\\ \hline
graph-mst & $n$, size of random graph to build \\ \hline
graph-bfs & $n$, size of random graph to build \\ \hline
graph-pagerank & $n$, size of random graph to build\\ \hline
dynamic-html & $n$, random number to generate an HTML page\\ \hline
chameleon & $\{n, m\}$, number of rows and columns to create an HTML table \\ \hline
\end{tabular}%
}
\end{table}

\begin{figure*}[t]
    \centering
    \begin{subfigure}{0.22\textwidth}
        \includegraphics[width=\textwidth]{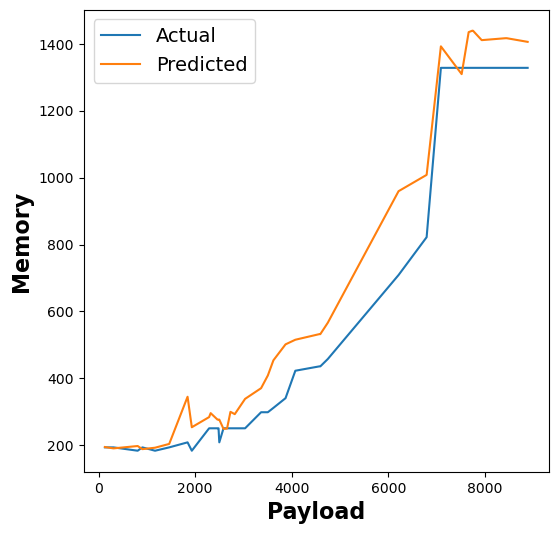}
        \caption{RFR Memory utilisation estimates for \textit{linpack}}
        \label{fig:linpack_memory_rfr}
    \end{subfigure}
    \hfill
        \begin{subfigure}{0.22\textwidth}
        \includegraphics[width=\textwidth]{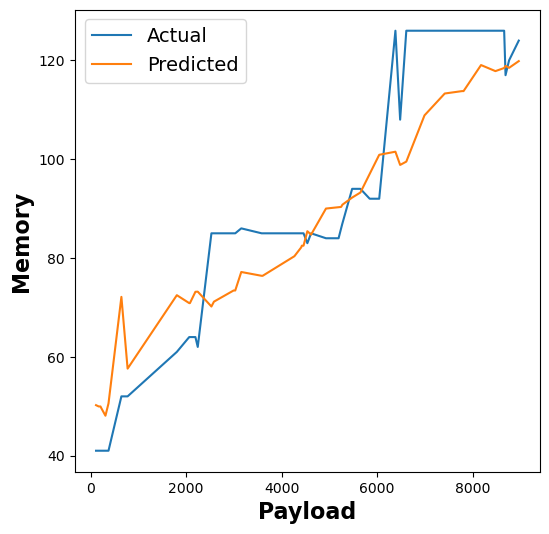}
        \caption{RFR Memory utilisation estimates for \textit{graph-pagerank}}
        \label{fig:pagerank_memory_rfr}
    \end{subfigure}
        \hfill
    \begin{subfigure}{0.22\textwidth}
        \includegraphics[width=\textwidth]{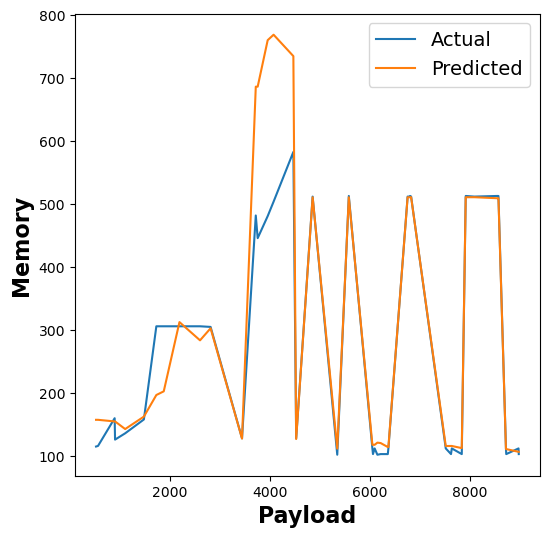}
        \caption{RFR Memory utilisation estimates for \textit{matmul}}
        \label{fig:matmul_memory_rfr}
    \end{subfigure}
    \hfill
    \begin{subfigure}{0.22\textwidth}
        \includegraphics[width=\textwidth]{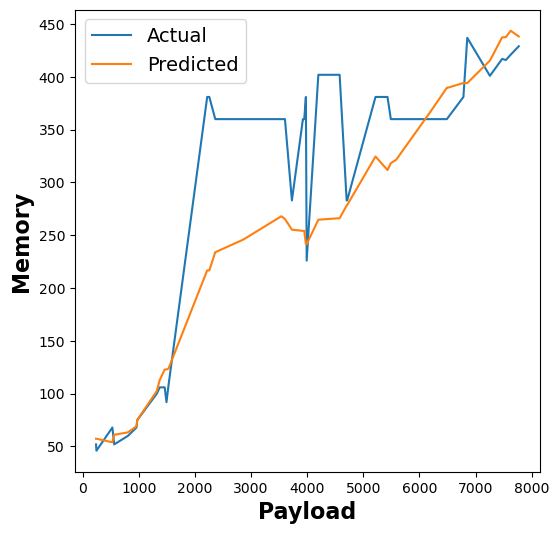}
        \caption{RFR Memory utilisation estimates for \textit{graph-bfs}}
        \label{fig:bfs_memory_rfr}
    \end{subfigure}
    \caption{RFR Model Payload-Aware Memory utilisation Prediction.}
    \label{fig:rfr_memory}
\end{figure*}

\subsection{Experiments}
\label{subsec:experiments}

We run the profiling step of the proposed solution with the given function payload(s) at distinct memory configurations to capture the relationship between payload and resource demands. After the profiling step, a RFR model is trained on the collected data, accompanied by a hyper-parameter tuning that explores model parameters such as \textit{n\_estimators, max\_depth, min\_samples\_split and min\_samples\_leaf} via \textit{grid\_search} and selects the best configuration for inference. The RFR model then estimates the memory utilisation and function execution time to perform online optimisation and selects the best possible payload-aware configuration to invoke functions.

We perform the payload-aware estimation and optimisation of memory configuration for $50$ payload values, within the discussed range. We select the relative importance, $w_z$, as $0.5$ to balance the function execution time, $T_f(m,P)$ and run-time cost, $C_f(m,P)$ constraints for this experiment. In Fig. \ref{fig:rfr_prediction}, we showcase the ability of RFR model to predict the execution time of the functions. The proposed methodology estimated the execution time with a $R^2$ score of as high as $98$\% for \textit{linpack} and \textit{pyaes} function while having $R^2$ scores of $97$\%, $94$\% and $91$\% for functions such as \textit{graph-pagerank, graph-mst, graph-bfs} and \textit{dynamic-html}. This statistical measure of $R^2$ score demonstrates the goodness of the fit by the regression model. Additionally, we observe in Fig. \ref{fig:rfr_memory} that for memory-intensive functions, the RFR model is able to capture the relationship of payload and memory with high $R^2$ score of $96$\% in case of \textit{linpack}, $87$\% for \textit{graph-pagerank}, $79$\% for \textit{matmul} and as low as $73$\% for \textit{graph-bfs} function with a mean absolute error (MAE) of $269$ MB, $36$MB, $2609$ MB and $192$ MB, respectively. These observations are based on the actual performance data acquired after invoking functions with RFR estimated values. Therefore, we can conclude from the observations that RFR model can be utilised for predicting payload-aware function execution time and memory utilisation. In addition, the proposed RFR-based framework can take advantage of online estimation and optimised solutions to invoke the respective payload-aware configurations.

To evaluate our model's efficiency in reducing excess resource allocation and higher run-time costs, we compare our work with the following existing works - 

\begin{enumerate*}
    \item COSE \cite{COSE.V2}: a Bayesian Optimisation (BO) based function memory configuration tool that tries to select best configuration at each sample which maximises the model confidence. 
    \item Parrotfish \cite{ParrotFish}: an online Parametric Regression based function configuration tool that selects optimal configuration while satisfying user-defined constraints. 
    \item AWS Lambda Power Tuning \cite{AWSPowerTuning}: a recommendation and graphical tool by AWS that performs an exhaustive search of memory configurations to suggest a cheaper and lower execution time configuration.
\end{enumerate*}

\begin{figure*}[t]
    \centering
    \begin{subfigure}{0.22\textwidth}
        \includegraphics[width=\textwidth]{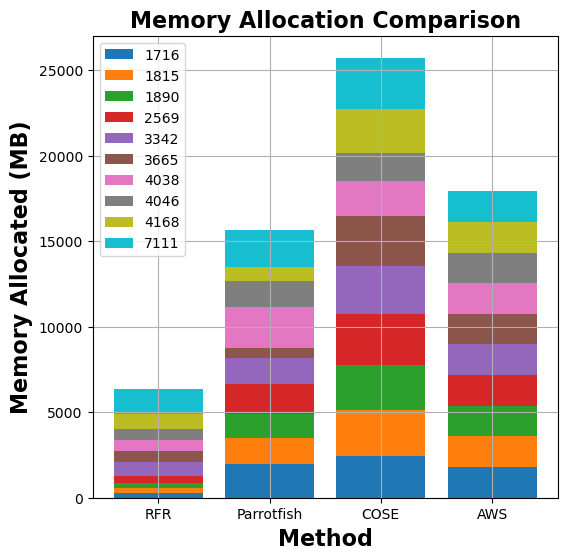}
        \caption{Comparison of memory allocation for \textit{graph-mst}}
        \label{fig:mst_memory}
    \end{subfigure}
    \hfill
    \begin{subfigure}{0.22\textwidth}
        \includegraphics[width=\textwidth]{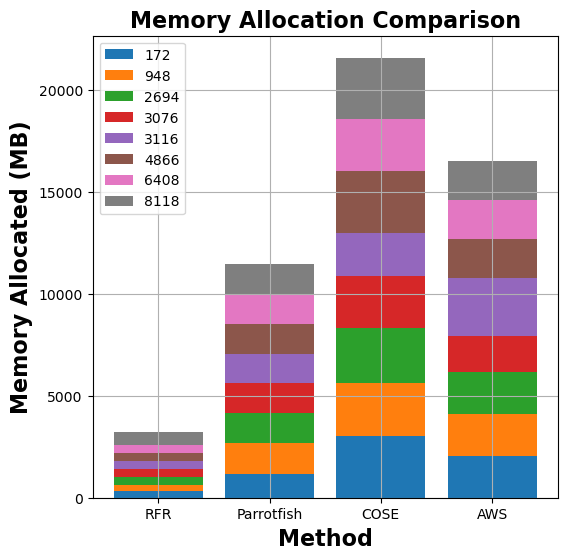}
        \caption{Comparison of memory allocation for \textit{pyaes}}
        \label{fig:pyaes_memory}
    \end{subfigure}
    \hfill
        \begin{subfigure}{0.22\textwidth}
        \includegraphics[width=\textwidth]{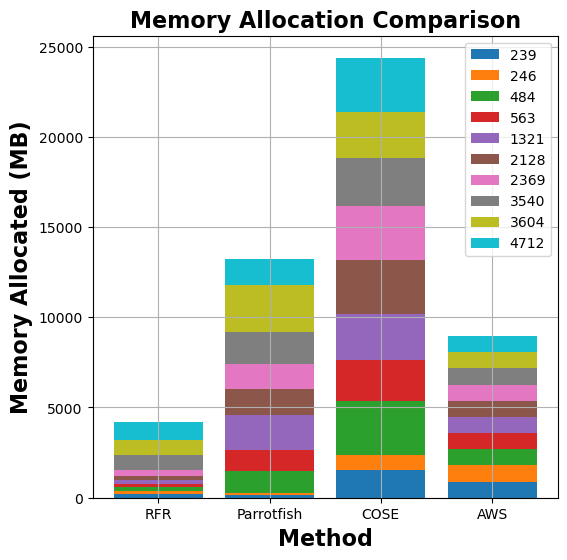}
        \caption{Comparison of memory allocation for  \textit{graph-bfs}}
        \label{fig:bfs_memory}
    \end{subfigure}
    \hfill
    \begin{subfigure}{0.22\textwidth}
        \includegraphics[width=\textwidth]{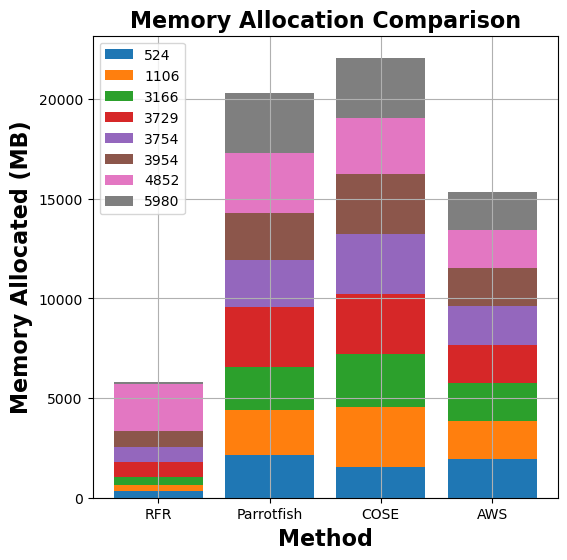}
        \caption{RFR Execution time estimates for \textit{matmul}}
        \label{fig:matmul_memory}
    \end{subfigure}
    \vskip\baselineskip
    \begin{subfigure}{0.22\textwidth}
        \includegraphics[width=\textwidth]{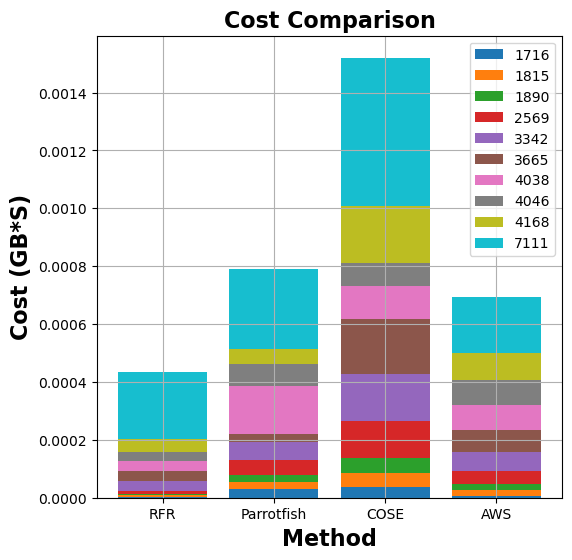}
        \caption{Comparison of run-time cost for \textit{graph-mst}}
        \label{fig:mst_cost}
    \end{subfigure}
    \hfill
        \begin{subfigure}{0.22\textwidth}
        \includegraphics[width=\textwidth]{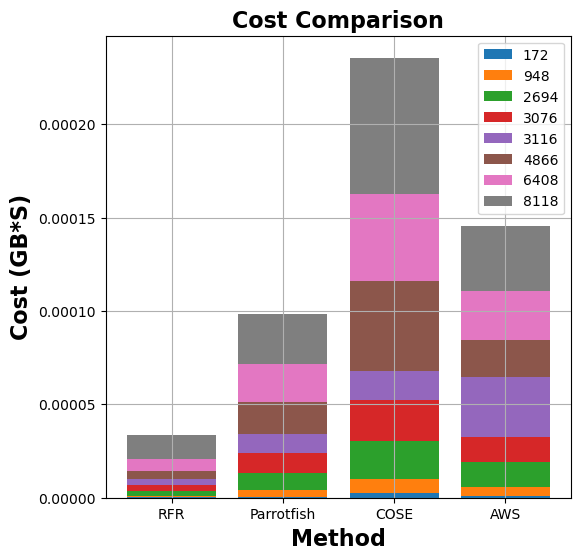}
        \caption{Comparison of run-time cost for \textit{pyaes}}
        \label{fig:pyaes_cost}
    \end{subfigure}
    \hfill
        \begin{subfigure}{0.22\textwidth}
        \includegraphics[width=\textwidth]{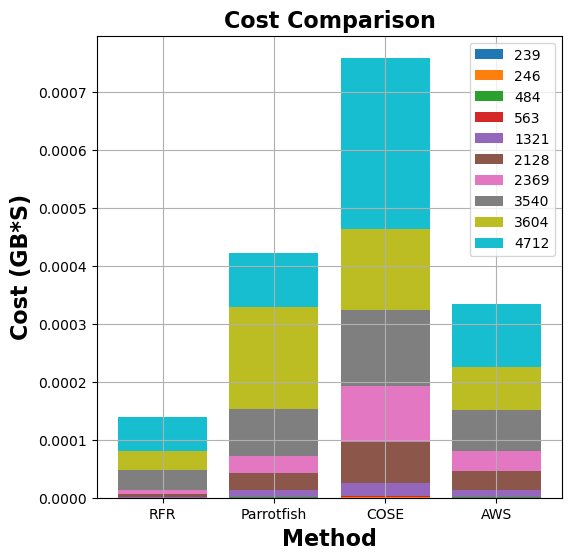}
        \caption{Comparison of run-time cost for \textit{graph-bfs}}
        \label{fig:bfs_cost}
    \end{subfigure}
        \hfill
    \begin{subfigure}{0.22\textwidth}
        \includegraphics[width=\textwidth]{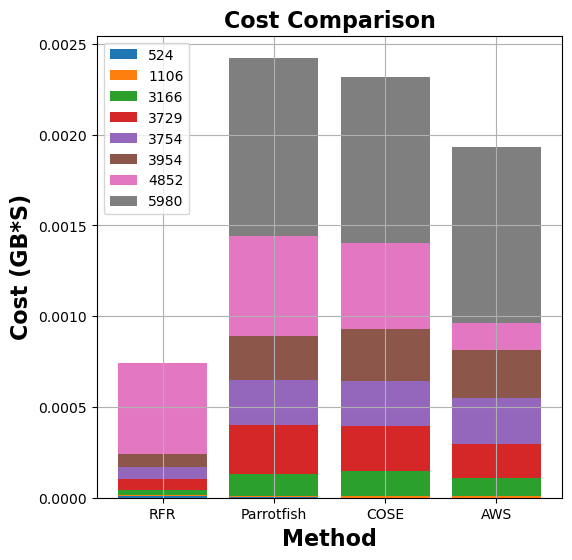}
        \caption{Comparison of run-time cost for \textit{matmul}}
        \label{fig:matmul_cost}
    \end{subfigure}
    
    \caption{Comparison of Memory Allocation and Run-time Costs for competing approaches}
    \label{fig:evaluation}
\end{figure*}

For the brevity of this evaluation, we run the respective approaches for atmost $10$ payload values to find the payload-aware optimal memory configuration for $4$ functions i.e., \textit{graph-mst, pyaes, matmul} and \textit{graph-bfs}. Similar results are reported for other functions and thus have been skipped from this discussion. We select the function execution time as the objective and utilise distinct payloads to predict the execution time. However, COSE does not provide any utility or provision to optimise for specific payloads, therefore, we run the COSE tool to probe $20$ sample points for each payload value. On the other hand, Parrotfish samples and tries to optimise the memory configuration based on weighted representative payloads via parametric regression. The tool recommends the individual optimal memory configuration found i.e., the cheapest configuration within the deadline, for that specific run with the weighted payload(s). However, we run Parrotfish for each distinct payload to find the optimal configuration and ignore any interference effect of other payloads. We also run \cite{AWSPowerTuning} for individual payloads with a set of memory configurations that do not raise any runtime errors. 

In Fig. \ref{fig:evaluation} we share the results of MemFigLess estimation and execution as compared to COSE and Parrotfish. The results are optimised for $1.5$ times the function deadline, $D$ and no weight is given to the run-time cost. However, we observe that the proposed RFR-based MemFigLess is able to estimate and utilise the Pareto optimal results to allocate a lower memory configuration as compared to other works, given a function deadline. In terms of memory allocation, MemFigLess allocates $54$\%, $75$\% and $65\%$ less cumulative memory as compared to Parrotfish, COSE and AWS Power Tuning for \textit{graph-mst}. Additionally, this allocation allows to save $57$\%, $79$\% and $58\%$ in cumulative run-time costs of \textit{graph-mst} against when run with Parrotfish and COSE selected configuration. The gains are more visible for \textit{pyaes} function, where MemFigLess is able to save $82$\% additional resources as compared to COSE leading to $84$\% cost benefits. Similar results are achieved for \textit{graph-bfs} where MemFigLess saved $65$\% and $75\%$ resources as compared to Parrotfish and AWS Power Tuning, and was $87$\% cost efficient in comparison to COSE. For \textit{matmul} function, the resource savings are approximately $73$\% as compared to COSE and Parrotfish. 
Therefore, we conclude based on the experimental analysis that the RFR-based solution is able to reduce the run-time costs and excess resource allocation as compared to SOTA techniques, \cite{COSE.V2}\cite{ParrotFish}\cite{AWSPowerTuning}, while satisfying the deadline constraint.

\subsection{Discussion}
\label{subsec:discussion}

We investigate a payload-aware function configuration methodology and present MemFigLess, a RFR-based workflow to estimate the payload-aware optimal resource allocation schemes. 
The experiments are performed with distinct functions deployed on the AWS Lambda platform and take advantage of workflows to create the offline profiling and training stages of MemFigLess. We assume that representative payload(s) are provided for profiling to supplement the regression model. In addition to this, we make a heuristics-based decision to allocate maximum memory, $3008$ MB to an unseen payload larger than the profiled limit or to configure the memory of the smallest payload seen. This builds on the idea \cite{ParrotFish} that if a configuration is good for an input $x$, then it is also good, if not better, for inputs smaller than $x$.

We profile and train the functions at memory intervals of $128$ MB, however, in the online inference, estimates are made for every possible memory configuration with a step size of $1$ MB, in line with \cite{ParrotFish} and \cite{Lambda}. This makes the inference time complexity $O(n + k \log k)$  where $n$ represents the number of explored memory configurations during optimisation with $k \leq n$ configurations in Pareto front calculation. As we employ a Random Forest regression that generally requires prior data for precise estimates, these finer estimates may suffer in case of complex payload and resource relationships. Therefore, the accuracy of the estimates is highly dependent on the memory intervals and representative payloads used in the initial profiling to capture complex resource relationships. Furthermore, finer online inferences may suffer from increased cold starts if the estimated configurations are largely distinct for incoming payloads. To support this, MemFigLess intelligently checks for existing functions with estimated configuration to execute. However, it does not control the degree of container reuse that utilises a warm function configuration to avoid a cold start. In addition to this, a sequential workload is considered for analysis in anticipation of minor performance fluctuations owing to AWS Lambda guaranteed concurrent invocations. With the discussed experimental assumptions and setup, the proposed solution is able to reduce the excess resource allocation and reduce the run-time cost of functions in comparison to Parrotfish and COSE, which are used as is with their defaults.

\section{Conclusions and Future Work}
In this work, we present MemFigLess, a Random Forest regression-based payload-aware solution to optimise function memory configuration. This solution is implemented using AWS serverless services and deployed as a workflow. A motivation study is conducted to highlight the importance of payload-aware resource configuration for performance guarantees. We identify a strong and positive correlation between function payload, execution time and memory configuration to formalise the resource configuration as a multi-objective optimisation problem. A concept of Pareto dominance is utilised to perform online resource optimisation. We compare the proposed solution to COSE and Parrotfish and demonstrate the effectiveness of RFR in reducing resource wastage and saving costs. MemFigLess is able to reduce excess memory allocation of as high as $85$\% against COSE and save run-time cost of up to $71$\% contrasting to Parrotfish.

As part of future work, we plan to improve MemFigLess by exploring advanced optimisation techniques such as Evolutionary algorithms and meta-heuristics to reduce the sampling costs. Additionally, we aim to explore the integration of Reinforcement Learning algorithms and LLMs to reduce manual efforts in model refinement. Furthermore, as Random Forest regression training requires training data prior to learning, an incremental learning approach can also be employed. Furthermore, an integration of Distributed and Federated learning techniques can also be explored to address faster and larger training scenarios.

\bibliography{citation}

\begin{thebibliography}{10}

\bibitem{CNCF}
CNCF, ``Cloud native computing foundation - serverless landscape.'' \url{https://landscape.cncf.io/serverless}, 2023.

\bibitem{BerkleyView2019}
E.~Jonas, J.~Schleier-Smith, V.~Sreekanti, C.-C. Tsai, A.~Khandelwal, Q.~Pu, V.~Shankar, J.~Carreira, K.~Krauth, N.~Yadwadkar, {\em et~al.}, ``Cloud programming simplified: A berkeley view on serverless computing,'' {\em arXiv preprint arXiv:1902.03383}, 2019.

\bibitem{Lambda}
AWS, ``Aws lambda - run code without thinking about servers or clusters.'' \url{https://aws.amazon.com/lambda/}, 2023.

\bibitem{GCF}
Google, ``Cloud functions.'' \url{https://cloud.google.com/functions}, 2023.

\bibitem{Azure}
Microsoft, ``Azure functions.'' \url{https://learn.microsoft.com/en-us/azure/azure-functions/}, 2023.

\bibitem{OpenFaaS}
OpenFaaS, ``Openfaas - serverless functions made simple.'' \url{https://docs.openfaas.com/}, 2021.

\bibitem{Knative}
Knative, ``Knative is an open-source enterprise-level solution to build serverless and event driven applications.'' \url{https://knative.dev/docs/}, 2022.

\bibitem{OpenWhisk}
OpenWhisk, ``Apache openwhisk - open source serverless cloud platform.'' \url{https://openwhisk.apache.org/}, 2016.

\bibitem{ServerlessVideo}
AWS, ``Serverlessvideo: Connect with users around the world!.'' \url{https://serverlessland.com/}, 2023.

\bibitem{NetflixAWS}
AWS, ``Serverless case study - netflix.'' \url{https://dashbird.io/blog/serverless-case-study-netflix/}, 2020.

\bibitem{CapitalOne}
CapitalOne, ``Capital one saves developer time and reduces costs by going serverless on aws.'' \url{https://aws.amazon.com/solutions/case-studies/capital-one-lambda-ecs-case-study/}, 2023.

\bibitem{AI-ML-AWS}
E.~Johnson, ``Deploying ml models with serverless templates.'' \url{https://aws.amazon.com/blogs/compute/deploying-machine-learning-models-with-serverless-templates/}, 2021.

\bibitem{LLM-Medium-AWS}
A.~Sojasingarayar, ``Build and deploy llm application in aws.'' \url{https://medium.com/@abonia/build-and-deploy-llm-application-in-aws-cca46c662749}, 2024.

\bibitem{Sizeless}
S.~Eismann, L.~Bui, J.~Grohmann, C.~Abad, N.~Herbst, and S.~Kounev, ``Sizeless: Predicting the optimal size of serverless functions,'' in {\em Proceedings of the 22nd International Middleware Conference}, (New York, NY, USA), p.~248–259, 2021.

\bibitem{PredictingTheCostsOfWorkflows}
S.~Eismann, J.~Grohmann, E.~van Eyk, N.~Herbst, and S.~Kounev, ``Predicting the costs of serverless workflows,'' in {\em Proceedings of the ACM/SPEC International Conference on Performance Engineering}, ICPE '20, (New York, NY, USA), p.~265–276, 2020.

\bibitem{StepConf}
Z.~Wen, Y.~Wang, and F.~Liu, ``Stepconf: Slo-aware dynamic resource configuration for serverless function workflows,'' in {\em Proceedings of IEEE Conference on Computer Communications}, pp.~1868--1877, 2022.

\bibitem{WithGreatFreedomComesGreatOpportunity}
M.~Bilal, M.~Canini, R.~Fonseca, and R.~Rodrigues, ``With great freedom comes great opportunity: Rethinking resource allocation for serverless functions,'' in {\em Proceedings of the 18th European Conference on Computer Systems}, (New York, NY, USA), p.~381–397, 2023.

\bibitem{COSE.V2}
A.~Raza, N.~Akhtar, V.~Isahagian, I.~Matta, and L.~Huang, ``Configuration and placement of serverless applications using statistical learning,'' {\em in IEEE Transactions on Network and Service Management}, vol.~20, no.~2, pp.~1065--1077, 2023.

\bibitem{CPUTAMS}
R.~Cordingly, S.~Xu, and W.~Lloyd, ``Function memory optimization for heterogeneous serverless platforms with cpu time accounting,'' in {\em IEEE International Conference on Cloud Engineering}, pp.~104--115, 2022.

\bibitem{Datadog2020}
Datadog, ``The state of serverless, 2020.'' \url{https://www.datadoghq.com/state-of-serverless-2020/}, 2020.

\bibitem{OFC}
D.~Mvondo, M.~Bacou, K.~Nguetchouang, L.~Ngale, S.~Pouget, J.~Kouam, R.~Lachaize, J.~Hwang, T.~Wood, D.~Hagimont, N.~De~Palma, B.~Batchakui, and A.~Tchana, ``Ofc: An opportunistic caching system for faas platforms,'' in {\em Proceedings of the 16th European Conference on Computer Systems}, (New York, NY, USA), p.~228–244, 2021.

\bibitem{Cypress}
V.~M. Bhasi, J.~R. Gunasekaran, A.~Sharma, M.~T. Kandemir, and C.~Das, ``Cypress: Input size-sensitive container provisioning and request scheduling for serverless platforms,'' in {\em Proceedings of 13th Symposium on Cloud Computing}, (New York, NY, USA), p.~257–272, 2022.

\bibitem{ParrotFish}
A.~Moghimi, J.~Hattori, A.~Li, M.~Ben~Chikha, and M.~Shahrad, ``Parrotfish: Parametric regression for optimizing serverless functions,'' in {\em Proceedings of the 2023 ACM Symposium on Cloud Computing}, (New York, NY, USA), p.~177–192, 2023.

\bibitem{FuncitonBench}
J.~Kim and K.~Lee, ``Functionbench: A suite of workloads for serverless cloud function service,'' in {\em Proceedings of the 12th International Conference on Cloud Computing}, pp.~502--504, 2019.

\bibitem{CloudWatch}
AWS, ``Amazon cloudwatch.'' \url{https://docs.aws.amazon.com/cloudwatch/}, 2024.

\bibitem{Lachesis}
P.~Sinha, K.~Kaffes, and N.~J. Yadwadkar, ``Online learning for right-sizing serverless functions,'' in {\em Proceedings of Architecture and System Support for Transformer Models}, 2023.

\bibitem{MAFF}
T.~Zubko, A.~Jindal, M.~Chadha, and M.~Gerndt, ``Maff: Self-adaptive memory optimization for serverless functions,'' in {\em Service-Oriented and Cloud Computing}, pp.~137--154, Springer International Publishing, 2022.

\bibitem{Astrea}
J.~Jarachanthan, L.~Chen, F.~Xu, and B.~Li, ``Astrea: Auto-serverless analytics towards cost-efficiency and qos-awareness,'' {\em Proceedings of IEEE Transactions on Parallel and Distributed Systems}, vol.~33, no.~12, pp.~3833--3849, 2022.

\bibitem{SLAM}
G.~Safaryan, A.~Jindal, M.~Chadha, and M.~Gerndt, ``Slam: Slo-aware memory optimization for serverless applications,'' in {\em Proceedings of the 15th International Conference on Cloud Computing}, pp.~30--39, 2022.

\bibitem{TimeCostMemoryConfig}
Z.~Li, H.~Yu, and G.~Fan, ``Time-cost efficient memory configuration for serverless workflow applications,'' {\em Proceedings of Concurrency and Computation: Practice and Experience}, vol.~34, no.~27, p.~e7308, 2022.

\bibitem{ModellingandOptimisationOfPerformanceCost}
C.~Lin and H.~Khazaei, ``Modeling and optimization of performance and cost of serverless applications,'' {\em Proceedings of IEEE Transactions on Parallel and Distributed Systems}, vol.~32, no.~3, pp.~615--632, 2020.

\bibitem{AWSPowerTuning}
A.~Casalboni, ``Profiling functions with aws lambda power tuning.'' \url{https://github.com/alexcasalboni/aws-lambda-power-tuning}, 2023.

\bibitem{RandomForest}
L.~Breiman, ``Random forests,'' {\em Machine learning}, vol.~45, pp.~5--32, 2001.

\bibitem{RFCloudCPU}
R.~B. Uriarte, S.~Tsaftaris, and F.~Tiezzi, ``Service clustering for autonomic clouds using random forest,'' in {\em Proceedings of 15th IEEE/ACM International Symposium on Cluster, Cloud and Grid Computing}, pp.~515--524, 2015.

\bibitem{RFAutoscaler23}
L.~M. Al~Qassem, T.~Stouraitis, E.~Damiani, and I.~A.~M. Elfadel, ``Proactive random-forest autoscaler for microservice resource allocation,'' {\em IEEE Access}, vol.~11, pp.~2570--2585, 2023.

\bibitem{RFCompVision}
A.~Saffari, C.~Leistner, J.~Santner, M.~Godec, and H.~Bischof, ``On-line random forests,'' in {\em 2009 IEEE 12th International Conference on Computer Vision Workshops, ICCV Workshops}, pp.~1393--1400, 2009.

\bibitem{AIBlock_Golec}
M.~Golec, D.~Chowdhury, S.~Jaglan, S.~S. Gill, and S.~Uhlig, ``Aiblock: Blockchain based lightweight framework for serverless computing using ai,'' in {\em Proceedings of 22nd IEEE International Symposium on Cluster, Cloud and Internet Computing (CCGrid)}, pp.~886--892, 2022.

\bibitem{ParetoMOO}
P.~Ngatchou, A.~Zarei, and A.~El-Sharkawi, ``Pareto multi objective optimization,'' in {\em Proceedings of the 13th International Conference on, Intelligent Systems Application to Power Systems}, pp.~84--91, 2005.

\bibitem{AWS_Serverless_Suite}
AWS, ``Serverless on aws.'' \url{https://aws.amazon.com/serverless/}, 2024.

\bibitem{scikit-learn}
F.~Pedregosa, G.~Varoquaux, A.~Gramfort, V.~Michel, B.~Thirion, O.~Grisel, M.~Blondel, P.~Prettenhofer, R.~Weiss, V.~Dubourg, J.~Vanderplas, A.~Passos, D.~Cournapeau, M.~Brucher, M.~Perrot, and E.~Duchesnay, ``Scikit-learn: Machine learning in {P}ython,'' {\em Journal of Machine Learning Research}, vol.~12, pp.~2825--2830, 2011.

\bibitem{SebsBenchmark}
M.~Copik, G.~Kwasniewski, M.~Besta, M.~Podstawski, and T.~Hoefler, ``Sebs: A serverless benchmark suite for function-as-a-service computing,'' in {\em Proceedings of the 22nd International Middleware Conference}, Middleware '21, (New York, NY, USA), p.~64–78, 2021.

\end{thebibliography}
\bibliographystyle{ieeetr}

\end{document}